\documentclass[preprint,groupedaddress,nofootinbib,preprintnumbers,eqsecnum]{revtex4-1}

\usepackage{amsmath} 

\usepackage{bm}

\begin{document}

\title{Gauge Freedom in  Path Integrals in Abelian Gauge Theory} 

\author{Teijiro Saito}
\affiliation{Graduate School of Science and Engineering, Yamagata University, 
Yamagata 990-8560, Japan 
}
\author{Ryusuke Endo}
\email{endo@sci.kj.yamagata-u.ac.jp}
\affiliation{Department of Physics, Yamagata University, Yamagata, 990-8560, Japan}
\author{Hikaru Miura}
\affiliation{Department of Physics, Yamagata University, Yamagata, 990-8560, Japan}

\begin{abstract}
We extend  gauge symmetry of Abelian gauge field to incorporate 
quantum  gauge degrees of freedom. 
We twice apply the Harada--Tsutsui gauge recovery procedure to 
gauge-fixed theories.  
First, starting from the Faddeev--Popov path integral in the Landau gauge, 
we recover the gauge symmetry by introducing an additional field as an 
extended gauge degree of freedom. Fixing the extended gauge symmetry by the 
usual Faddeev--Popov procedure, we 
obtain the theory of Type I gaugeon formalism.  
Next, applying the same procedure to the resulting gauge-fixed theory, 
we obtain a theory equivalent to the extended Type I gaugeon formalism. 

\end{abstract}

\preprint{YGHP-15-04}

\maketitle

\section{Introduction}

The standard formalism of canonically quantized gauge 
theories \cite{Nakanishi,Kugo78a,Kugo78b,Kugo79,Kugo89} 
does not consider quantum-level gauge transformations. 
There is no quantum gauge freedom, 
since the quantum theory is defined only after the gauge fixing.
Within the broader framework of Yokoyama's gaugeon 
formalism \cite{Yokoyama74a,Yokoyama74b,Yokoyama75a,Yokoyama75b,
Yokoyama78a,Yokoyama78b,Yokoyama78c,Yokoyama78d,Yokoyama78e,Yokoyama80}, 
we  can consider quantum gauge transformations as $q$-number gauge transformations 
among a family of Lorentz covariant linear gauges. 
In this formalism, quantum gauge freedom is provided 
by a set of extra fields, called gaugeon fields. 
The gaugeon formalism has been studied  not only in  Abelian 
fields \cite{Yokoyama74a,Yokoyama74b,Yokoyama78a,Izawa,Koseki,Endo} and Yang-Mills 
fields \cite{Yokoyama78b,Yokoyama78c,Yokoyama78d,Yokoyama78e,Yokoyama80,Abe,Koseki96,Upadhyay14} 
but also in the Higgs model \cite{Yokoyama75a,Miura,Upadhyay14}, 
chiral gauge theory \cite{Yokoyama75b}, Schwinger's model \cite{Schwinger}, 
 Rarita-Schwinger field \cite{Endo00}, string theory \cite{Faizal12a,Faizal12b}, 
and gravity \cite{Upadhyay14a,Upadhyay14b}.

Yokoyama and Kubo \cite{Yokoyama74b} proposed two types of gaugeon theories 
for  Abelian gauge fields, 
which they referred to as 
Type I 
and Type II theories. 
The Lagrangian of each theory has a gauge fixing 
parameter $\alpha$ that 
can be shifted 
from $\alpha$ to $\alpha$ + $\tau$ 
by a $q$-number gauge 
transformation. 
The tree level photon propagator can be expressed as
\begin{equation}
           \langle A_\mu A_\nu \rangle 
           \sim 
           \frac{1}{k^2}\left( g_{\mu\nu} + (a-1)\frac{k_\mu k_\nu}{k^2} \right),
\label{eq:tree_level_propagator}
\end{equation} 
where the parameter $a$ is defined as
\begin{subequations}
\begin{align}
    a &= \varepsilon\alpha^2   \quad (\varepsilon = \pm 1) \quad \text{for Type  I}, 
\label{1.2a}
\\
%
a&= \alpha        \qquad  \text{\hspace{4.5em} for  Type  II}.
\end{align}
\end{subequations}
In Type I theory, the $q$-number gauge transformation can change the absolute value, 
but not the sign,  of the parameter $a$; 
in Type II theory the parameter $a$ can be arbitrarily altered. 

The Lagrangian of the Abelian gauge field $A_\mu$ in Type I theory \cite{Yokoyama74a, Koseki} 
is given by
\begin{equation}
    {\mathcal L} 
        = -\frac{1}{4}F_{\mu\nu}F^{\mu\nu} 
         - \partial_\mu B A^\mu 
         - \partial_\mu Y_\ast \partial^\mu Y 
         + \frac{\varepsilon}{2}(Y_\ast + \alpha B)^2 
         - i\partial_\mu{c_\ast}\partial^\mu c 
         - i\partial_\mu K_\ast\partial^\mu K,
         \label{Eq.1a}
\end{equation}
where $F_{\mu\nu} = \partial_\mu A_\nu -\partial_\nu A_\mu$, 
$B$ is the Nakanishi--Lautrup field, ${ c_\ast}$ and $c$ 
are the usual Faddeev--Popov (FP) ghosts, $\alpha$ is the gauge fixing parameter, 
$Y$ and $Y_\ast$ are gaugeon fields, and $K$ and $K_\ast$ are FP ghosts 
for the gaugeon fields, which are introduced to ensure the BRST symmetry \cite{Koseki}. 
This Lagrangian permits the $q$-number gauge transformation 
where we vary the gauge fixing parameter $\alpha$. 
The transformation is defined by
\begin{equation}
\begin{split}
        A_\mu &\rightarrow {\hat A_\mu} = A_\mu + \tau\partial_\mu Y,  
        \\
        Y_\ast &\rightarrow {\hat Y_\ast} = Y_\ast - \tau B,  
        \\
        B &\rightarrow {\hat B} = B, \quad Y \rightarrow {\hat Y} = Y,  
        \\
        c &\rightarrow {\hat c} = c + \tau K, 
                       \quad c_\ast \rightarrow {\hat c_\ast} = c_\ast,  
        \\
        K &\rightarrow {\hat K} = K, 
                       \quad K_\ast \rightarrow {\hat K_\ast} = K_\ast -\tau c_\ast,  
\end{split}
\end{equation}
with $\tau$ being a parameter of the transformation. 
Under this transformation the Lagrangian (\ref{Eq.1a}) 
becomes 
\begin{equation}
                {\mathcal L}(\phi_A; \alpha) = {\mathcal L}({\hat \phi}_A; {\hat \alpha}),
\end{equation}
where $\phi_A$ collectively represents all fields  and {$\hat \alpha$} 
is defined by
\begin{equation}
{\hat \alpha} = \alpha + \tau.
\end{equation}

The Lagrangian and  $q$-number gauge transformation in Type II theory 
are described in \cite{Yokoyama74b}. BRST symmetric Type II theory is 
given by 
\cite{Izawa}. 

The Lagrangian of the 
extended Type I theory, 
investigated  by Endo \cite{Endo}, 
is given by
\begin{equation}
\begin{split}
       {\mathcal L} = &-\frac{1}{4}F_{\mu\nu}F^{\mu\nu} 
                       - \partial_\mu B A^\mu 
                       + (Y_{1\ast} + \alpha_1 B)(Y_{2\ast} + \alpha_2 B) 
                       - \partial_\mu Y_{1\ast}\partial^\mu Y_1 
                       - \partial_\mu Y_{2\ast}\partial^\mu Y_2 
                       \\
                      &- i\partial_\mu{c_\ast}\partial^\mu c 
                       - i\partial_\mu K_{1\ast}\partial^\mu K_1 
                       - i\partial_\mu K_{2\ast}\partial^\mu K_2,
                 \label{Eq.1b}
\end{split}
\end{equation} 
where $Y_i$ and $Y_{i\ast}\ (i =1,  2)$ are two sets of gaugeon fields, 
$K_i$ and $K_{i\ast}$ are two sets of FP ghosts for the gaugeon fields, 
and the constant ${\alpha_i}$ are the  gauge fixing parameters. 
The corresponding parameter $a$ of the tree level photon propagator 
(\ref{eq:tree_level_propagator})
 is given 
by $a = 2\alpha_1 \alpha_2$. 
Thus, this theory extends the Type I gaugeon formalism by setting 
$a$ as quadratic in the gauge fixing parameters (cf. (\ref{1.2a})).
Because the parameter $a$ (and its sign) can be changed into 
arbitrary values by the $q$-number gauge transformation 
($
            \alpha_1 \rightarrow\ \alpha_1 + \tau_1, 
            \ 
            \alpha_2 \rightarrow \alpha_2 + \tau_2
       $), 
this theory possesses some characteristics of Type II theory. 
The Lagrangian can also be written as 
\begin{equation}
\begin{split}
     {\mathcal  L} = &
          -\frac{1}{4}F^{\mu\nu}F_{\mu\nu} 
          - \partial_\mu B A^\mu 
          + \frac{1}{2}(Y_{+\ast} + \alpha_+ B)^2 
          - \frac{1}{2}(Y_{-\ast} + \alpha_- B)^2 
          - \partial_\mu Y_{+\ast}\partial^\mu Y_{+} 
          \\
          &
          - \partial_\mu Y_{-\ast}\partial^\mu Y_{-} 
           - i\partial_\mu c_\ast\partial^\mu c 
           - i\partial_\mu K_{+\ast}\partial^\mu K_+ 
           - i\partial_\mu K_{-\ast}\partial^\mu K_-.    
\end{split}
\label{Eq.1c}
\end{equation}
where $Y_{\pm}$ are defined by
\begin{equation}
        Y_{\pm} = \frac{1}{\sqrt 2}(Y_1 \pm Y_2)
\label{Eq.1d}
\end{equation}
and $Y_{\pm\ast}, K_\pm, K_{\pm\ast}$, and $\alpha_\pm$ are defined similarly. 
    
Gaugeon theories for the Yang--Mills fields have been proposed by 
Yokoyama  \cite{Yokoyama78b} 
and 
Yokoyama, Takeda and Monda \cite{Yokoyama80}. 
The BRST symmetric theories have  been obtained by Abe \cite{Abe} and 
Koseki, Sato and Endo \cite{Koseki96}. 
Although these theories are  easily shown 
to be equivalent to 
the standard formalism in the Landau gauge ($a$ = 0),  
their equivalence to the standard formalism in non-Landau gauges ($a \ne 0$) 
cannot be demonstrated. 
Therefore, these theories should be compared with the Abelian gaugeon theory, 
which is equivalent 
to non-Landau gauge theory ($a \ne 0$) 
as well as to the Landau gauge ($a = 0$) \cite{Koseki}.

Sakoda \cite{Sakoda} extended the gauge freedom of Yang--Mills fields 
using the gauge recovery procedure for gauge non-invariant functionals proposed 
by Babelon, Schaposnik and Viallet \cite{Babelon} 
and 
Harada and Tsutsui \cite{Harada,Harada87}. 
Sakoda's theory includes the two gauges of the standard formalism: 
the Landau gauge and a non-Landau $a$-gauge. 
Sakoda's theory considers the total Fock space, which embeds the Fock spaces 
of the both gauges of the standard formalism. 
In this theory, 
the  $q$-number gauge transformation connects the Landau gauge 
and non-Landau $a$-gauge. 
Different from 
the gaugeon formalism, the $q$-number transformation of Sakoda's theory 
cannot arbitrarily change the gauge parameter, 
but allows only 
$\alpha = 0$ and $\alpha = a$. 

In this paper, we further extend the gauge freedom 
to allow more flexibility in the gauge parameter than in Sakoda's theory. 
As a first step,  we consider 
the Abelian gauge field. 
Starting with the Faddeev--Popov path integral in the Landau gauge,  
we extend the gauge freedom by twice applying  the Harada--Tsutsui 
gauge recovery procedure \cite{Harada87}. In contrast, 
Sakoda \cite{Sakoda} applied this procedure once to the Yang--Mills field.

The remainder of this paper is organized as follows. Section  2 reviews 
the Harada--Tsutsui 
gauge recovery procedure for  gauge non-invariant functionals \cite{Harada87}  
and Sakoda's path integral \cite{Sakoda} of Yang--Mills fields. 
In Section 3, we extend the gauge symmetry of Abelian gauge fields 
twice using the Harada--Tsutsui gauge recovery procedure. 
In Section 4, we relate 
our theory to the gaugeon formalism  
and show that our theory is equivalent to the extended Type I gaugeon formalism.

\section{Path integral of the gauge non-invariant functional}
\subsection{Harada--Tsutsui gauge recovery procedure}
Harada and Tsutsui's procedure extends  the gauge degrees-of-freedom of the gauge 
non-invariant functional \cite{Harada87}. 
We illustrate their procedure on a system of 
gauge non-invariant Yang--Mills fields $A_\mu$. Such a system might comprise
massive Yang--Mills fields. 

The action $S_0[A]$ of the system is not invariant
\begin{equation}
       S_0[A^g] \neq S_0[A],
\end{equation}
under the gauge transformation,
\begin{equation}
       A_\mu  \rightarrow  A_\mu^g = gA_\mu g^{-1} + ig\partial_\mu g^{-1},
\end{equation}
where $g$ is a group-valued function. 
The usual path integral of the system is given by 
\begin{equation}
     Z_0 = \int {\mathcal D}A_\mu \,e^{iS_0[A]},
\end{equation}
which 
leads to  non-renormalizable propagators in the massive Yang--Mills case. 

Now, we promote the group-valued function $g(x)$ to a dynamical variable, and define 
an extended action by 
\begin{equation}
     S[A, g] \equiv S_0[A^g],
\end{equation}
which is now invariant under the  extended gauge transformation,
\begin{equation}
\begin{array}{l}
       A \rightarrow A^h,  
       \\
       g \rightarrow g^h = gh^{-1},                                                          %
\end{array}
\end{equation}
where $h(x)$ is a group-valued function. 
The formal path integral for $S[A, g]$
\begin{equation}
      Z_{\rm div} = \int {\mathcal D}A{\mathcal D}g\,e^{iS[A, g]}
      \label{Eq. 2a}
\end{equation}
is divergent since $S[A, g]$ is gauge invariant. 
To factor out the divergent gauge volume, we require gauge fixing 
(if $g$ = 1, $Z_{\rm div}$ reduces to $Z_0$).   
Expressing the gauge fixing condition as
\begin{equation}
           f[g, A] =0,
           \label{eq.(2.7)}
\end{equation}
the corresponding FP determinant $\Delta_{\rm FP}[A, g]$ is given by 
\begin{equation}
              1 = \Delta_{\rm FP}[A, g]
                  \int {\mathcal D}h\delta(f[gh, A^{h^{-1}}]).
              \label{Eq. 2b}
\end{equation}
Inserting (\ref{Eq. 2b}) into (\ref{Eq. 2a}) and factoring out the gauge volume, 
we obtain 
\begin{equation}
          Z = \int{\mathcal D}A{\mathcal D}g\,\Delta_{\rm FP}[A, g]
                 \delta(f[g, A])\,e^{iS[A, g]}.
\end{equation}

We can also consider 't~Hooft averaging. 
Instead of the gauge fixing condition 
(\ref{eq.(2.7)}), 
we use
\begin{equation}
         f[g, A] = C(x),
\end{equation}
where $C(x)$ is an arbitrary $c$-number function. Averaging the path integral
over $C(x)$ with the Gaussian weight
\begin{equation}
        \exp\left[-\frac{i}{2a}\int d^4 x \, C(x)^2\right],
\end{equation}
we obtain 
\begin{equation}
\begin{split}
     Z &
        = \int {\mathcal D}A {\mathcal D}g {\mathcal D}\Phi {\mathcal D}C
          \Delta_{\rm FP}[A, g] 
               \,e^{ iS[A,g] 
                   + i\int d^4 x \{\Phi (f[g, A] - C) - {C^2}/{2a}\}}, 
        \\
      &
        =\int {\mathcal D}A {\mathcal D}g {\mathcal D}\Phi
          \Delta_{\rm FP}[A, g]
                   \,e^{iS[A, g] 
                     + i\int d^4 x\{ \Phi f[g, A] + {a}\Phi^2/2 \} }.
\end{split}
\label{PI2.1:Harada-Tsutui}
\end{equation}
The first line expresses the delta functional as a Fourier integral 
with respect to a field $\Phi$. 
Equation (\ref{PI2.1:Harada-Tsutui}) yields renormalizable propagators
in the massive Yang--Mills case.

\subsection{Sakoda's method}
Sakoda \cite{Sakoda} extended the gauge freedom
of the gauge-fixed Yang--Mills fields in the Landau gauge
by applying  the Harada--Tsutsui procedure. 
This method is briefly explained below. 
  
The Landau-gauge Lagrangian of a Yang-Mills field $A_\mu$ is given by
\begin{equation} 
          {\mathcal L}_{\rm L} = 2{\rm tr}\left[
                     - \frac{1}{4}F^{\mu\nu}F_{\mu\nu} 
                     + B\partial^\mu A_\mu 
                     + i{\bar c}\partial^\mu D_\mu c
                                          \right],
\end{equation}
where $F^{\mu\nu}$ is the field strength, 
$B$ is the Nakanishi--Lautrup field, and $c$ and ${\bar c}$ are the FP ghosts.
We express the path integral as 
\begin{gather}
\begin{split}
           Z_0 &= \int{\mathcal D}A{\mathcal D}B{\mathcal D}{\bar c}{\mathcal D}c \,
                       e^{i\int {\mathcal L}_{\rm L}d^4 x}
           \\
             &= \int {\mathcal D}A{\mathcal D}B\,I_0[A, B],  
\end{split}
\end{gather}
where 
\begin{equation}
       I_0[A, B] = \Delta[A]
                 e^{
                    i\int {d^4 x 2{\rm tr} 
                          (-\frac{1}{4}F^{\mu\nu} F_{\mu\nu} + B\partial^\mu A_\mu)}
                   }, 
\end{equation}
\begin{equation}
         \Delta[A] = \det  \partial^\mu D_\mu. 
\end{equation}
Since we consider a gauge fixed system,
the functional $I_0[A, B]$ is not gauge invariant under the gauge transformation
\begin{equation}
\begin{array}{l}
            A_\mu \rightarrow A_\mu^g = gA_\mu g^{-1} + ig\partial_\mu g^{-1},  
            \\ 
            B \rightarrow B^g = B. 
\end{array}
\end{equation}
Now, we promote the group-valued function $g(x)$ to a dynamical variable 
and 
define
\begin{equation}
     \tilde I_0[A, B, g] \equiv I_0[A^g, B^g] 
                  = \Delta[A^g] \, \exp \left\{
                                 i\int d^4 x \, 2{\rm tr} \Bigl(-\frac{1}{4}F^{\mu\nu} F_{\mu\nu} 
                                    + B\partial^\mu A_\mu^g \Bigr)
                                    \right\}. 
\end{equation}
The functional $\tilde I_0[A, B, g]$ 
is invariant under the extended gauge transformation,
\begin{equation}
\begin{array}{l}
          A \rightarrow A^h,  
          \\
          g \rightarrow g^h = gh^{-1},  
          \\
          B \rightarrow B^h=B,
\end{array}
\end{equation}
where $h(x)$ is a group-valued function. 
The formal path integral for $\tilde I_0[A, B, g],$ 
\begin{equation}
           Z_{\rm div}  
              = \int {\mathcal D}A{\mathcal D}B{\mathcal D}g\, \tilde I_0[A, B, g],
          \label{Eq. 2c}
\end{equation}
is divergent since $\tilde I_0[A, B, g]$ is now gauge invariant. 
To factor out the divergent gauge volume,  we require 
gauge fixing.   
For this purpose, we consider the following gauge fixing condition
\begin{equation}
          f[g, A] \equiv \partial^\mu A_\mu - \partial^\mu A_\mu^g = C,
\end{equation}
where $C$ is an arbitrary $c$-number function. 
The corresponding FP determinant $\Delta_{\rm FP}[A, g, C]$ 
is then given by
\begin{equation}
1 = \Delta_{\rm FP}[A, g, C]\int {\mathcal D}h\delta(f[gh, {A^h}^{-1}] - C).
\label{Eq. 2d}
\end{equation}
Inserting (\ref{Eq. 2d}) into (\ref{Eq. 2c}) and factoring out the gauge volume,
we obtain
\begin{equation}
            Z = \int {\mathcal D}A{\mathcal D}B{\mathcal D}g
                            \,\tilde I_0[A, B, g] \Delta[A] \delta(f[g, A] - C),
           \label{Eq. 2e}
\end{equation}
In (\ref{Eq. 2e}),  $\Delta_{\rm FP}[A, g, C]$ was evaluated as 
\begin{equation}
\begin{split}
       \Delta_{\rm FP}[A, g, C] \delta(f[g, A] - C) 
               &= \det (\partial^\mu D_\mu) \delta(f[g, A] - C) 
               \\
               & = \Delta[A]\delta(f[g, A] - C). 
\end{split}
\end{equation}
Expressing the delta functional as a Fourier integral 
with respect to $\Phi$ and applying 't~Hooft averaging 
with a  Gaussian weight, we obtain
\begin{equation}
              Z = \int {\mathcal D}A{\mathcal D}B{\mathcal D}g{\mathcal D}\Phi\, 
                     \tilde I_0[A, B, g] \Delta[A] 
                     \exp \left\{
                          i\int d^4x\, 2{\rm tr}\Bigl( \Phi f[g, A] + \frac{a}{2}\Phi^2 \Bigr)
                         \right\},
\end{equation}
where $a$ is the gauge fixing parameter.
The corresponding Lagrangian is given by
\begin{equation}
          {\mathcal L} = 2{\rm tr}\left[
                            - \frac{1}{4}F^{\mu\nu}F_{\mu\nu} 
                            + B\partial^\mu A_\mu^g 
                            + i{\bar \eta}\partial^\mu D_\mu^g \eta^g 
                            + \Phi f[g, A]  
                            + \frac{a}{2}\Phi^2 
                            + i{\bar c}\partial^\mu D_\mu c
                                  \right].
          \label{Eq.2f}
\end{equation}
Here, the determinant $\Delta[A^g]$ 
in $\tilde I_0[A, B, g]$ has been expressed 
in terms of the 
FP ghosts $\eta$ and ${\bar \eta}$: 
\begin{equation}
             \Delta[A^g] = \int {\mathcal D}{\bar \eta}{\mathcal D}\eta\, 
                        \exp \Bigl\{
                                 i\int d^4 x\,2{\rm tr}\bigl(
                                          i{\bar \eta}\partial^\mu D_\mu^g\eta^g
                                                       \bigr)
                             \Bigl\},
\end{equation} 
where
\begin{equation}
                  D_\mu^g \eta^g =\partial^\mu \eta^g - i[A_\mu^g, \eta^g], 
                    \qquad 
                  \eta^g = g\eta g^{-1}.
\end{equation}

The Lagrangian (\ref{Eq.2f}) is invariant under the following BRST transformations,  
$\bm{\delta}$, 
$\tilde {\bm{\delta}}$, 
and 
$\bm{\delta}_{\rm B} = \bm{\delta} + \tilde {\bm{\delta}}$,
\begin{equation}
\begin{split}
          &\bm{\delta}A_\mu = D_\mu c,   \qquad  \bm{\delta}g = -igc, 
          \\
          &\bm{\delta}c = ic^2,   \qquad   \bm{\delta}\eta = i\{c, \eta\},
          \\
          &\bm{\delta}{\bar c} = i\Phi,   
             \qquad  \bm{\delta}\Phi = \bm{\delta}B = \bm{\delta}{\bar \eta} = 0,
\label{Eq.2g}
\end{split}
\end{equation}
and
\begin{equation}
\begin{split}
          &{\tilde {\bm{\delta}}}A_\mu = 0, \qquad
                    {\tilde {\bm{\delta}}}g = -ig\eta, 
          \\
          &{\tilde {\bm{\delta}}}\eta = i\eta^2, \qquad
                      {\tilde {\bm{\delta}}}c = 0,  \qquad
                              {\tilde {\bm{\delta}}}{\bar c} = 0,
          \\
          &{\tilde {\bm{\delta}}}{\bar \eta} = i(\Phi - B),  \qquad
                 {\tilde {\bm{\delta}}}\Phi = {\tilde {\bm{\delta}}}B = 0. 
\end{split}
\label{Eq.2h}
\end{equation}
These transformations satisfy the nilpotency condition, 
$
        {\bm{\delta}}^2 = {\bm{\tilde \delta}}^2 
                       = {\bm{\delta}_{\rm B}}^2 
                       = \{\bm{\delta}, \tilde {\bm{\delta}}\} 
                       = 0. 
$
We denote 
the corresponding BRST charges by $Q, {\tilde Q}$, and $Q_{\rm B}$.

In Sakoda's theory, we can consider the two subspaces of the total Fock space 
using the BRST charges.
One is the subspace 
$\ker \tilde Q = \{\vert\phi\rangle;\, \tilde Q \vert \phi \rangle = 0\}$;  
the other is  $\ker Q$.  
The subspace $\ker \tilde Q$ corresponds to the Fock space of 
the standard formalism of the $a$-gauge. 
To see this,  we express the Lagrangian (\ref{Eq.2f}) 
as\footnote{%
   We comment here that another expression 
   \begin{align*}
    {\mathcal L} 
        = &2 \mathrm{tr} \big[
            - \frac{1}{4}F^{\mu\nu}F_{\mu\nu} 
            + \Phi \partial^\mu A_\mu + \frac{a}{2}\Phi^2 
            + i{\bar c}\partial^\mu D_\mu c
                 \,\big] 
            + i\tilde {\bm{\delta}} \left( 2 \mathrm{ tr} \big [
                     {\bar \eta} \,\partial^\mu A_\mu^g
                     \,\big] \right),
     \tag{\ref{eq:Sakodas_a-gaugeExpression}'}
     \end{align*}
     would be 
     helpful to analyze the subspace $\ker \tilde Q$ by using the BRST charge $Q$.
     The field $\Phi$ 
     (rather than $B$) 
     plays the role of Nakanishi--Lautrup field in this $a$-gauge theory. 
   }
\begin{align}
    {\mathcal L} 
        = &2 \mathrm{tr} \big[
            - \frac{1}{4}F^{\mu\nu}F_{\mu\nu} 
            + B\partial^\mu A_\mu + \frac{a}{2}B^2 
            + i{\bar c}\partial^\mu D_\mu c
                 \,\big] 
          \notag \\
          &  - i\tilde {\bm{\delta}} \left( 2 \mathrm{ tr} \big [
                     {\bar \eta}\big \{(\partial^\mu A_\mu - \partial^\mu A_\mu^g) 
                     + \frac{a}{2}(\Phi + B) \big \}
                     \,\big] \right).
\label{eq:Sakodas_a-gaugeExpression}
\end{align}
The term in the second line is $\tilde Q$-exact, and thus ignorable 
in the subspace $\ker \tilde Q$; the remaining term is nothing but the $a$-gauge Lagrangian. 
To show that the subspace $\ker Q$ corresponds to the Fock space of the standard formalism 
of the Landau gauge, we denote the  
Landau-gauge 
fields $A_\mu^\prime$ by $A_\mu^\prime = A_\mu^g$ and express the Lagrangian as
\begin{align}
    {\mathcal L} = &2 \mathrm {tr} \big[
                      -\frac{1}{4}F^{\prime\mu\nu}F_{\mu\nu}^\prime 
                      + B\partial^\mu A_\mu^\prime 
                      + i{\bar \eta}\partial^\mu D_\mu^\prime\eta^\prime
                      \big ] 
                 \notag
                 \\
                 &-i \bm{\delta} \left( 2\mathrm {tr} \big [
                      {\bar c}\big \{ 
                           (\partial ^\mu{A_\mu^\prime}^{g^{-1}} - \partial^\mu A_\mu^\prime) 
                           + \frac{a}{2}\Phi
                             \big \}
                                 \big ]\right),
\end{align}
where $F_{\mu\nu}^\prime$ is the field strength of $A_\mu^\prime$, $D_\mu^\prime$ 
is the covariant derivative corresponding to $A_\mu^\prime$, and $\eta^\prime=\eta^g$.
The term in the second line is ignorable in the subspace $\ker Q$; 
the remaining term is the Landau-gauge Lagrangian. 
Thus, in Sakoda's theory, the subspaces of the total Fock space 
identify the Fock spaces of the  standard theory of the $a$-gauge and Landau gauge. 
The  $a$-gauge field $A_\mu$ and  Landau-gauge field $A_\mu^\prime$ 
are connected through the  $q$-number gauge transformation $g(x)$. 
The $q$-number transformation of Sakoda's theory limits the gauge parameter 
to  only two values, $\alpha = 0$ and $\alpha = a$. Considering this, 
Sakoda's theory differs from the gaugeon 
formulation of the Yang--Mills field \cite{Yokoyama80, Abe, Koseki96}. 
(Strictly speaking, 
using Sakoda's $q$-number transformation 
$g(x)$ 
we can define another $q$-number gauge transformation $\{g(x)\}^{\tau}$
with an arbitrary real number $\tau$. 
This transformation changes the gauge parameter $a$ 
into $a(1-\tau)^2$  
in the tree level propagator of $A_\mu$.  
We do not, however, consider 
this transformation at present, 
since the transformed Lagrangian would have complicated 
terms and it would not be easy to analyze the theory in this gauge.%
\footnote{%
         In  the Abelian limit, the situation becomes simple. 
         The Abelian limit of  Sakoda's theory is 
         equivalent to the Abelian gaugeon formalism 
         (see sections \ref{subsectionSakoda} and \ref{subsection2nd}); 
         the transformation $\{g(x)\}^\tau$ becomes a usual $q$-number 
         gauge transformation of the Abelian gaugeon formalism.  
        } )

\section{Successive extension of the gauge freedom of the Abelian field}
\subsection{Sakoda's extension}\label{subsectionSakoda}
For the selfcontainedness of this section, 
we repeat here Sakoda's arguments in the Abelian case. 

We start with
the Landau-gauge Lagrangian of the Abelian gauge field given by 
\begin{equation}
      {\mathcal L}_{\rm L} = 
              -\frac{1}{4}F^{\mu\nu}F_{\mu\nu} 
                   + B\partial^\mu A_\mu 
                      + i{\bar c}\partial^\mu\partial_\mu c.
\end{equation}
The path integral is expressed as
\begin{equation}
\begin{split}
            Z_0 &= \int {\mathcal D}A{\mathcal D}B{\mathcal D}{\bar c}{\mathcal D}c
               \,e^{i\int d^4 x{\mathcal L}_{\rm L}} 
               \\
              &= \int {\mathcal D}A{\mathcal D}B \,I_0[A, B], 
\end{split}
\end{equation}
where
\begin{align}
         I_0[A, B] &= \Delta \, 
              e^{i\int d^4x(-\frac{1}{4}F^{\mu\nu}F_{\mu\nu} + B\partial^\mu A_\mu)},
          \\
          \Delta &= \det  \partial^\mu\partial_\mu.
\end{align}
Since we consider a gauge fixed system,
the functional $I_0[A, B]$ is not gauge invariant under the gauge transformation
\begin{equation}
\begin{array}{l}
        A_\mu \rightarrow A_\mu^\theta = A_\mu + \partial_\mu \theta, 
        \\
        B \rightarrow B^\theta = B,
\end{array}
\end{equation} 
where $\theta$ is an arbitrary scalar function.
Now, we promote the function $\theta$ to a dynamical variable and 
define
\begin{equation}
%
      \tilde I_0[A, B, \theta] \equiv I_0[A^\theta, B^\theta]
                     = \Delta \,e^{i\int d^4x\{-\frac{1}{4}F^{\mu\nu}F_{\mu\nu} 
                       + B\partial^\mu(A_\mu + \partial_\mu\theta)\}}.
%
\end{equation}
The functional $\tilde I_0[A, B, \theta]$ 
is invariant under the extended gauge transformation,
\begin{equation}
\begin{array}{l}
        A_\mu \rightarrow A_\mu + \partial_\mu \lambda,  
        \\
        \theta \rightarrow \theta  - \lambda,  
        \\
        B \rightarrow B,
\end{array}
\end{equation}
where $\lambda$ is an arbitrary scalar function.
The formal path integral for $\tilde I_0[A, B, \theta]$,
\begin{equation}
     Z_{{\rm div}} = \int {\mathcal D}A {\mathcal D}B {\mathcal D}\theta  \,
                       \tilde I_0[A, B, \theta],
\label{Eq. 3a}
\end{equation}
is divergent since $\tilde I_0[A, B, \theta]$ is now gauge invariant. 
To factor out the divergent gauge volume, we require gauge fixing.
For this purpose, we consider the following gauge fixing condition
\begin{equation}
        f[\theta, A] \equiv \partial^\mu A_\mu - \partial^\mu A_\mu^\theta = C. 
\end{equation}
The corresponding FP determinant is then given by
\begin{equation}
1 = \Delta_{\rm FP}[A, \theta, C] \int {\mathcal D}\lambda
                      \delta(-\partial^\mu\partial_\mu\theta^\lambda - C).
\label{Eq. 3b}
\end{equation}
Inserting (\ref{Eq. 3b}) into (\ref{Eq. 3a}) 
and factoring out the gauge volume, we obtain 
\begin{equation}
        Z_1 = \int {\mathcal D}A {\mathcal D}B {\mathcal D}\theta \,
                  \tilde I_0[A, B, \theta] \Delta\delta(-\partial^\mu\partial_\mu\theta - C),
\label{Eq. 3c}
\end{equation}
where we have evaluated  $\Delta_{\rm FP}[A, \theta, C]$ as
%
\begin{align}
       \Delta_{\rm FP}[A, \theta, C] \, \delta(-\partial^\mu\partial_\mu\theta - C) 
          &= \det (\partial^\mu\partial_\mu) \, \delta(-\partial^\mu\partial_\mu\theta - C)  
          \notag \\
          & = \Delta \, \delta(-\partial^\mu\partial_\mu\theta - C).
\end{align}
Expressing the delta functional as a Fourier integral with respect to $\Phi$ 
and applying 't~Hooft averaging with a Gaussian weight, we obtain
\begin{equation}
    Z_1 = \int {\mathcal D}A {\mathcal D}B {\mathcal D}\theta {\mathcal D}\Phi \,
                         I_1[A, B, \theta, \Phi], 
\label{Eq.3d}
\end{equation}
with
\begin{equation}
\begin{split}
      I_1[A, B, \theta, \Phi] = \Delta \Delta 
                \exp \Bigl[
                   i\int d^4x\Bigl\{
                           - \frac{1}{4}F^{\mu\nu}F_{\mu\nu} 
                           + B\partial^\mu(A_\mu + \partial_\mu\theta) 
                           - \Phi \partial^\mu\partial_\mu \theta 
                           +\frac{a}{2}\Phi^2
                               \Bigr\}
                      \Bigr],
\end{split}
\end{equation}
where $a$ is the gauge fixing parameter.
The two FP determinants can be expressed in terms of two pairs of ghost fields
\begin{equation}
    \Delta \Delta = \int {\mathcal D}{\bar c} {\mathcal D}c 
                         {\mathcal D}{\bar \eta} {\mathcal D}\eta \, 
                         \exp \Bigl[ 
                             i\int d^4x \big\{ i{\bar c}\partial^\mu\partial_\mu c 
                                              + i{\bar \eta}\partial^\mu\partial_\mu \eta
                                        \big\}
                              \Bigr].   
\end{equation}
Summarizing these results,  
we obtain the Lagrangian of the first extension of the gauge freedom as 
\begin{equation}
    {\mathcal L}_{\rm 1st} = - \frac{1}{4}F^{\mu\nu}F_{\mu\nu} 
                   + B\partial^\mu(A_\mu + \partial_\mu \theta)
                   - \Phi \partial^\mu \partial_\mu \theta  
                   + \frac{1}{2}a \Phi^2 
                   + i\bar c\partial^\mu\partial_\mu c 
                   + i\bar \eta\partial^\mu\partial_\mu \eta.
\label{Eq.3e}
\end{equation}
Because we extended the gauge freedom by the method of Sakoda \cite{Sakoda}, 
the above Lagrangian is the Abelian limit of Sakoda's Yang--Mills 
Lagrangian (\ref{Eq.2f}). 
As implied by Sakoda \cite{Sakoda}, (\ref{Eq.3e}) is equivalent to the Lagrangian of the gaugeon 
formalism. 
(We will confirm this in section \ref{subsection2nd}.)
The Lagrangian is invariant under the following 
Abelian version of Sakoda's BRST transformation: 
\begin{equation}
\begin{split}
   &\bm{\delta}_{\rm B}A_\mu = \partial_\mu c, 
          \qquad  \bm{\delta}_{\rm B}\theta = -(c + \eta),
          \\
   &\bm{\delta}_{\rm B}{\bar c} = i\Phi,   
          \qquad \bm{\delta}_{\rm B}{\bar \eta} = i(\Phi -B),
          \\
   & \bm{\delta}_{\rm B}{B} = \bm{\delta}_{\rm B}\Phi 
              = \bm{\delta}_{\rm B}c = \bm{\delta}_{\rm B}\eta = 0.
\end{split}
\end{equation} 
This transformation satisfies the nilpotency $\bm{\delta}_{\rm B}^2 =0$. 
We can also find $\bm{\delta}$ and $\bm{\tilde \delta}$ transformations 
satisfying $\bm{\delta}_{\rm B}$ = $\bm{\delta} + \bm{\tilde \delta}$, as 
in the Yang--Mills case (\ref{Eq.2g}) and (\ref{Eq.2h}).

\subsection{The successive extension}
Starting from the path integral (\ref{Eq.3d}), 
we again extend  the gauge freedom of the Lagrangian ${\mathcal L}_{\rm 1st}$. 
The functional $I_1[A, B, \theta, \Phi]$ 
is not gauge invariant under the gauge transformation,
\begin{equation}
\begin{array}{l}
    A_\mu \rightarrow A_\mu^{\chi} = A_\mu + \partial_\mu {\chi},  
\\
    \theta \rightarrow \theta^{\chi} = \theta  - {\chi},
\\
    B \rightarrow B^{\chi} = B, 
\\
    \Phi \rightarrow \Phi^{\chi} = \Phi,                                                     %
\end{array}
\end{equation}
where ${\chi}$ is an arbitrary scalar function.
Now, we promote the function ${\chi}$ to a dynamical variable and define
\begin{equation}
\begin{split}
      \tilde I_1[A, &B, \theta, \Phi, {\chi}] 
          \equiv I_1[A^{{\chi}}, B^{{\chi}}, \theta^{{\chi}}, \Phi^{{\chi}}] 
          \\
          &= 
          \Delta \Delta\, \exp \biggl[i\int d^4 x 
                                        \Bigl\{
                               - \frac{1}{4}F^{\mu\nu}F_{\mu\nu}
                               + B\partial^\mu (A_\mu + \partial_\mu \theta) 
                               - \Phi\partial^\mu\partial_\mu (\theta - {\chi}) 
                               + \frac{1}{2}a \Phi^2
                                        \Bigr\}
                               \biggr],
\end{split}
\end{equation}
where we have used 
$A_\mu^{\chi} + \partial_\mu \theta^{\chi} = A_\mu + \partial_\mu\theta$. 
The functional $\tilde I_1[A, B, \theta, \Phi, {\chi}]$ is gauge invariant 
under the following extended gauge transformation:
\begin{equation}
\begin{array}{l}
 A_\mu  \rightarrow A_\mu + \partial_\mu \lambda,  
 \\ 
 \theta \rightarrow \theta - \lambda,  
 \\
 {\chi} \rightarrow {\chi} - \lambda,      
 \\
 B \rightarrow B, 
 \\
 \Phi  \rightarrow \Phi,
\end{array}
\end{equation}
where $\lambda$ is an arbitrary scalar function.
The formal path integral for $\tilde I_1[A, B, \theta, \Phi, {\chi}]$,
\begin{equation}
       Z_{\rm div} 
       = \int {\mathcal D}A{\mathcal D}B{\mathcal D}\theta {\mathcal D}\Phi 
         {\mathcal D}{\chi}\, 
         \tilde I_1[A, B, \theta, \Phi, {\chi}],
\label{Eq.3f}  
\end{equation}
is divergent since 
$\tilde I_1[A, B, \theta, \Phi, {\chi}]$ is now gauge invariant. 
To factor out the divergent gauge volume, we require gauge fixing. 
We consider the following gauge fixing condition
\begin{equation}
      f[{\chi},A] \equiv \partial^\mu A_\mu -\partial^\mu A_\mu^{{\chi}}  = C 
\end{equation}
where $C$ is an arbitrary $c$-number function. 
The corresponding FP determinant $\Delta_{\rm FP}[A, {\chi}, C]$ is then defined as 
\begin{equation}
     1 = \Delta_{\rm FP}[A, {\chi}, C]
        \int {\mathcal D}\lambda 
        \delta(-\partial^\mu\partial_\mu{\chi}^\lambda - C).
\label{Eq.3g}
\end{equation}
Inserting (\ref{Eq.3g}) into (\ref{Eq.3f}) 
and factoring out the gauge volume, we obtain
\begin{equation}
     Z_2 
        = \int {\mathcal D}A{\mathcal D}B{\mathcal D}\theta {\mathcal D}\Phi 
          {\mathcal D}{\chi} {\mathcal D}C\,
          \tilde I_1[A, B, \theta, \Phi, {\chi}]
          \Delta\delta(-\partial^\mu\partial_\mu {\chi} - C)
\end{equation}
where we have evaluated  $\Delta_{\rm FP}[A, {\chi}, C]$ as
\begin{align}
        \Delta_{\rm FP}[A, {\chi}, C] \,\delta(-\partial^\mu \partial_\mu -C)
            &= \det (\partial^\mu\partial_\mu)\delta(-\partial^\mu\partial_\mu {\chi} - C)
            \notag \\
            &= \Delta\delta(-\partial^\mu\partial_\mu {\chi} - C).
\end{align}
Expressing the delta functional as a Fourier integral 
with respect to $\phi$ and applying  't~Hooft averaging 
with a Gaussian weight, we obtain 
\begin{equation}
             Z_2 
                  = \int {\mathcal D}A{\mathcal D}B{\mathcal D}\theta {\mathcal D}\Phi 
                    {\mathcal D}{\chi} {\mathcal D}\phi 
                    \,I_2[A, B, \theta, \Phi, {\chi}, \phi], 
\end{equation}
with
\begin{equation}
\begin{split}
    I_2[A, B, \theta, \Phi, {\chi}, \phi] = \Delta\Delta\Delta{\rm exp}
              \biggl[&i\int d^4x\{-\frac{1}{4}F^{\mu\nu}F_{\mu\nu}
                                     + B\partial^\mu(A_\mu + \partial_\mu \theta)  
                    \\
                     &- \Phi\partial^\mu\partial_\mu(\theta -{\chi})
                                      + \frac{a}{2}\Phi^2 
                                       - \phi\partial^\mu\partial_\mu{\chi} 
                                       + \frac{a^\prime}{2}\phi^2\}\biggr],   
\end{split}
\end{equation}
where $a^\prime$ is another gauge fixing parameter.
The Lagrangian of the successive extension of the gauge degree of 
freedom  is expressed by
\begin{equation}
\begin{split}
          {\mathcal L_{\rm 2nd}} 
             = &-\frac{1}{4}F^{\mu\nu}F_{\mu\nu} + B\partial^\mu A_\mu 
                              + (B -\Phi)\partial^\mu\partial_\mu\theta
                              + (\Phi - \phi)\partial^\mu\partial_\mu{\chi}
                              + \frac{a}{2}\Phi^2 + \frac{a^\prime}{2}\phi^2 
         \\
               &+ i{\bar c}\partial^\mu\partial_\mu c
                             + i{\bar \eta}\partial^\mu\partial_\mu\eta 
                             + i{\bar \xi}\partial^\mu\partial_\mu\xi,
\label{Eq.3h}
\end{split}
\end{equation}
where the third $\Delta$ has been expressed in terms of the ghost fields $\xi$ and $\bar \xi$: 
\begin{equation}
    \det  (\partial^\mu\partial_\mu) 
         =\int {\mathcal D}{\bar \xi}{\mathcal D}\xi\, 
              \exp\left(i\int d^4x i{\bar \xi}\partial^\mu\partial_\mu\xi \right).      
\end{equation}
This Lagrangian is invariant under the following BRST transformation, 
\begin{equation}
\begin{split}
       &\bm{\delta}_{\rm B}A_\mu = \partial_\mu c ,   
               \qquad \qquad  \bm{\delta}_{\rm B}\theta = \eta,
       \\ 
       &\bm{\delta}_{\rm B}{\chi} = \xi,   
               \qquad \qquad  \bm{\delta}_{\rm B}{\bar c} = iB, 
       \\
       &\bm{\delta}_{\rm B}{\bar \eta} = i(B - \Phi),   
                \qquad \quad  \bm{\delta}_{\rm B}{\bar \xi} = i(\Phi - \phi), 
       \\
       &\bm{\delta}_{\rm B}{B} = \bm{\delta}_{\rm B}\Phi 
                 = \bm{\delta}_{\rm B}\phi  = \bm{\delta}_{\rm B}c 
                  = \bm{\delta}_{\rm B}{\eta} =  \bm{\delta}_{\rm B}\xi  = 0,
\end{split}
\end{equation} 
which  satisfies the nilpotency $\bm{\delta}_{\rm B}^2 =0.$

\section{Relation to the gaugeon formalism}
\label{section4}
\subsection{Equivalence of ${\mathcal L}_{\rm 1st}$ and the Type I gaugeon Lagrangian}%
\label{subsection2nd}

We first 
confirm that the Lagrangian (\ref{Eq.3e}) of Sakoda's extension 
is equivalent to the Lagrangian of Type I gaugeon theory (\ref{Eq.1a}).
Redefining the  fields as 
\begin{equation}
\begin{split}
       & \theta = -\alpha Y,  
        \\
       &\Phi = \frac{1}{\alpha}Y_\ast + B,  
        \\
       &\eta = K, \quad {\bar \eta} = K_\ast, 
\end{split}
\end{equation} 
where  $\alpha$ is a numerical parameter satisfying 
$a = \varepsilon\alpha^2$ ($\varepsilon = {a}/|a|$), 
we can rewrite the Lagrangian (\ref{Eq.3e})  as 
\begin{equation}
          {\mathcal L}_{\rm 1st}  
                    = -\frac{1}{4}F^{\mu\nu}F_{\mu\nu} + B\partial^\mu A_\mu 
                          + Y_\ast\partial^\mu\partial_\mu Y 
                          + \frac{1}{2}\varepsilon(Y_\ast + \alpha B)^2 
                          + ic_\ast\partial^\mu\partial_\mu c 
                          + iK_\ast\partial^\mu\partial_\mu K,
\end{equation}
which is exactly  (\ref{Eq.1a}).
It should be noted that the field $\theta$ introduced as an extended gauge freedom 
plays the role of a gaugeon field.

\subsection{Equivalence of $\mathcal{L}_{\rm 2nd}$ and the extended Type I Lagrangian}
Next, we  show that  the Lagrangian (\ref{Eq.3h}) of the 
successive extension is equivalent to  the extended Type I Lagrangian (\ref{Eq.1c}). 
Redefining the fields as
\begin{equation}
\begin{split}
      &\theta = -\alpha_+ Y_+ - \alpha_- Y_-,  
      \\
      &\Phi = \frac{1}{\alpha_+}Y_{+\ast} + B,  
      \\ 
      &\chi = -\alpha_- Y_-,  
      \\
      &\phi = \frac{1}{\alpha_-}Y_{-\ast} + B, 
      \\
      &\eta = K_+, \quad  {\bar \eta} = K_{+\ast},   
      \\
      &\xi = K_-, \quad   {\bar \xi} = K_{-\ast},
\end{split} 
\end{equation}
where  $\alpha_+$ and $\alpha_-$ are numerical parameters satisfying
\begin{equation}
     a = \varepsilon{\alpha_+}^2 \quad (\varepsilon = a/|a|), 
     \qquad \quad
     a^\prime = \varepsilon^\prime{\alpha_-}^2 \quad (\varepsilon^\prime = a'/|a'|),
\end{equation}
we can rewrite the Lagrangian (\ref{Eq.3h}) as
\begin{equation}
\begin{split}
       {\mathcal L}_{\rm 2nd} 
          = &-\frac{1}{4}F^{\mu\nu}F_{\mu\nu} 
                     + B\partial^\mu A_\mu 
                     + Y_{+\ast}\partial^\mu\partial_\mu Y_{+}  
                     + Y_{-\ast}\partial^\mu\partial_\mu Y_- 
                     + \frac{1}{2}\varepsilon(Y_{+\ast} + \alpha_+ B)^2
       \\
           &+ \frac{1}{2}\varepsilon^\prime(Y_{-\ast} + \alpha_- B)^2
                     + ic_\ast\partial^\mu\partial_\mu c 
                     + iK_{+\ast}\partial^\mu\partial_\mu K_+ 
                     + iK_{-\ast}\partial^\mu\partial_\mu K_-.
\end{split}
\label{Eq.4II}
\end{equation}
Provided that the sign factors $\varepsilon$ and $\varepsilon^\prime$ differ, 
this Lagrangian is equivalent to the extended Type I gaugeon 
Lagrangian (\ref{Eq.1c}). 
In the successive extension, 
the real scalar fields $\theta$ and $\chi$ introduced 
as extended gauge degrees of freedom 
play the roles of gaugeon fields $Y_+$ and $Y_-$.

\section{Summary and Discussion}
Starting from the Faddeev--Popov path integral 
in the Landau gauge of  Abelian gauge theory, 
we successively extended the gauge symmetry  to incorporate 
quantum gauge degrees of freedom. 
Following Sakoda's treatment of Yang--Mills fields, we applied 
(in each extension) 
the Harada--Tsutsui gauge recovery procedure to the gauge non-invariant functional.
The Lagrangian resulting from the first extension 
agrees with the Abelian limit of Sakoda's Yang--Mills Lagrangian,  
and is equivalent to that of Type I gaugeon theory. 
The scalar field $\theta$ introduced as an extended gauge degree of freedom 
plays the role of a gaugeon field $Y$. 
The theory obtained by the second extension is 
equivalent to extended Type I theory 
if the signs ($\varepsilon$ and $\varepsilon^\prime$) differ. 
The scalar fields $\theta$ and $\chi$ introduced as extended gauge degrees of freedom 
play the roles of gaugeon fields $Y_+$ and $Y_-$. 

One might think  what happens when we further repeat  Sakoda's extensions of gauge freedom. 
In the Abelian case, extending the gauge freedom three times or more 
would not yield  any new features,  
at least from  the viewpoint of the photon propagator (\ref{eq:tree_level_propagator}). 
For example, when we apply Sakoda's extension to the Lagrangian 
 ${\mathcal L}_{\rm 2nd}$   (\ref{Eq.4II}), 
we obtain the third pair of gaugeon fields $(Y_{3\ast},Y_3)$, 
corresponding FP ghosts $(K_{3\ast},K_3)$, and the third extended Lagrangian, 
\begin{align}
  {\mathcal L}_{\rm 3rd}
  = {\mathcal L}_{\rm 2nd}
  + \frac{1}{2}\varepsilon_3 (Y_{3\ast}+\alpha_3 B)^2 
  + Y_{3\ast}\partial^\mu \partial_\mu Y_3
  + iK_{3\ast}\partial^\mu \partial_\mu K_3,
\label{Eq.5III}
\end{align}
where 
$\varepsilon_3$ is a sign factor and $\alpha_3$  a numerical parameter. 
The photon propagator following from (\ref{Eq.5III}) 
is again expressed as 
(\ref{eq:tree_level_propagator}) where the parameter $a$ is now given by 
\begin{align}
  a=   \varepsilon (\alpha_+)^2 + \varepsilon^\prime (\alpha_-)^2
      + \varepsilon_3 (\alpha_3)^2. 
\end{align}
Thus the third extension does not expand the region of the values 
of the parameter $a$
; the second extension is enough to give the parameter $a$ 
an arbitrary value. 
In the non-Abelian  case, non-trivial features may appear 
when we apply Sakoda's extensions 
multiple times. Exploring this possibility  is our next task.


\end{document}